\documentstyle[prl,twocolumn,aps]{revtex}

\begin{document}

\draft

\title{Comment on ``Nature of the Spin Glass State''}  
\author{C.M.~Newman}
\address{Courant Inst.~of Math.~Sciences,
New York Univ., New York, NY 10012}
\author{D.L.~Stein}
\address{Departments of Physics and Mathematics, University of Arizona,
Tucson, AZ 85721}
\maketitle

\begin{abstract} 
We show that the interface excitations of Palassini-Young and
Krzakala-Martin cannot yield new thermodynamic states.
\end{abstract}



Recently, Palassini and Young \cite{PY00} numerically studied the ground
state structure of nearest-neighbor Ising spin glasses with Gaussian
couplings in spatial dimensions $d=3,4$.  Using a novel coupling-dependent
bulk term in the Hamiltonian, they looked for large-scale, low-energy
excitations above the ground state in fixed volume with periodic boundary
conditions (b.c.'s).  They interpret their results as evidence of spin
configurations whose interface with the ground state (i.e., couplings
satisfied in the excitation configuration but not in the ground state, or
vice-versa) scales as $L^{d_s}$, with $d_s < d$.  At the same time the
excitation energy was found to scale as $L^{\theta'}$, with $\theta'=0$;
i.e., it remained finite as $L \to \infty$ \cite{note}.  This is in
distinction to the similar exponent $\theta$, which governs the energy
difference under a change in coupling-independent b.c.'s, such as periodic
to antiperiodic.  In both $d=3,4$, numerical studies are consistent with
$\theta>0$ (see references in \cite{PY00}).  Similar results to \cite{PY00}
were found in $d=3$ by Krzakala and Martin
\cite{KM00}.

In this Comment, we show that the two properties $d_s<d$ and $\theta'=0$
are incompatible thermodynamically; i.e., such excitations cannot give rise
to new (ground or pure) states: if such interfaces exist, they cannot be
pinned, but rather must deflect to infinity as $L \to \infty$. (They could
also vanish altogether, as noted in \cite{PY00}.)

Suppose that these interfaces are pinned; this implies the existence
of infinite volume states that differ by interfaces with $d_s<d$ and whose
energy remains of order one as $L \to \infty$.  Consider two such states
$\alpha$ and $\beta$ and the periodic b.c.~metastate \cite{NS96b,NS97} for
the above model with periodic b.c.'s (our argument works also for other
b.c.'s chosen independently of the couplings); the metastate simply gives
the probability of finding a given ground state pair (GSP) in a typical
large volume.

As one moves along the interface, its energy should be larger sometimes in
$\alpha$ and sometimes in $\beta$, so in some volumes $\alpha$ would be the
ground state and in others $\beta$.  Both $\alpha$ and $\beta$ would
therefore have positive weight in the metastate, i.e., would appear in a
positive fraction of volumes.  For simplicity we assumed above only a
single excitation, but this argument can be extended to an arbitrary number
(even uncountably many).

But, as in \cite{NS00} (Lemma 2), if the periodic b.c.~metastate is
supported on multiple GSP's, these must have relative interfaces with
dimension $d_s=d$.  It follows that the conditions $d_s<d$ and $\theta'=0$
are incompatible for pinned interfaces on sufficiently large length scales.
(Pinned interfaces with $d_s<d$, leading to multiple ground states at zero
temperature and pure states at positive temperature, would probably require
$\theta'>(d-1)/2$ \cite{FH88,vE,NS92}, and could only be seen with
coupling-dependent b.c.'s.)

It follows that, if there exist excitations above the ground state with
$d_s<d$ and energy of order one independently of $L$, they would simply
deflect to infinity as $L\to\infty$.  So the possibility of their existence could be
relevant in determining the excitation spectrum of spin glasses, but they
would not imply the existence of new thermodynamic (ground or pure) states.

This research was supported in part by NSF Grants DMS-98-02310 (CMN) and
DMS-98-02153 (DLS).  The authors thank Peter Young and Matteo Palassini for
useful correspondence.

\end{document}